\documentclass[10pt,twocolumn]{article}
\usepackage[margin=0.8in]{geometry}
\usepackage{graphicx}
\usepackage{amsmath,amssymb}
\usepackage{authblk}
\usepackage{times}
\usepackage{caption}
\usepackage{float}

\title{\vspace{-0.5cm}Information Capacity of EEG: Theoretical and Computational Limits of Recoverable Neural Information\vspace{-0.3cm}}

\author[1]{Ishir Rao}
\affil[1]{Yale University}

\date{}

\begin{document}
\maketitle

\begin{abstract}
Electroencephalography (EEG) is widely used to study human brain dynamics, yet its quantitative information capacity remains unclear.
Here we combine information theory and synthetic forward modeling to estimate the mutual information between latent cortical sources and EEG recordings.
Using Gaussian-channel theory~\cite{cover,shannon} and empirical simulations, we find that scalp EEG conveys only tens of bits per sample about low-dimensional neural activity.
Information saturates with $\sim$64--128 electrodes and scales logarithmically with signal-to-noise ratio (SNR)~\cite{goldenholz}.
Linear decoders capture nearly all variance that is linearly recoverable, but the mutual information they recover remains far below the analytic channel capacity, indicating that measurement physics---not algorithmic complexity---is the dominant limitation.
These results outline the intrinsic ceiling on how much structure about brain state or thought content can be inferred from EEG.
\end{abstract}

\vspace{0.3cm}

\section*{1.\hspace{0.5em}Introduction}
EEG captures millivolt-scale fluctuations generated by large populations of cortical neurons.
Because these potentials are spatially blurred and noise-limited~\cite{nunez,grech}, the information they carry about the true cortical state is bounded.
We ask: \textit{What is the theoretical maximum mutual information between scalp recordings and underlying neural sources, and how close can practical decoders approach this limit?}

\begin{figure}[H]
\centering
\includegraphics[width=0.9\linewidth]{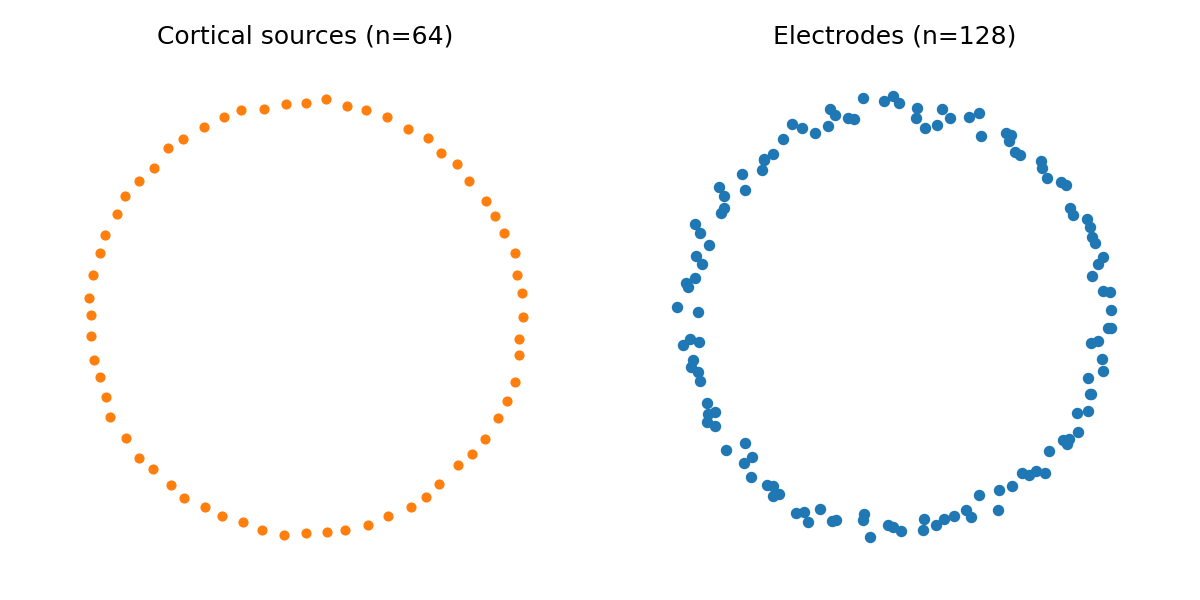}
\caption{Schematic of the synthetic forward model.
Cortical sources (orange) generate latent activity projected to scalp electrodes (blue) through a linear blur matrix $A$ with additive correlated noise $\varepsilon$.}
\label{fig:schematic}
\end{figure}

\section*{2.\hspace{0.5em}Model and Methods}
\textbf{Generative model.}
We simulate a two-dimensional cortical field of $n_s = 64$ sources arranged on a circle.
Each source’s activity arises from $n_\ell = 8$ latent processes following autoregressive (AR(1)) dynamics with coefficient $\rho = 0.9$.
Sensor readings are given by
\[
Y = A X + \varepsilon,
\]
where $A$ is a Gaussian spatial blur (leadfield) and $\varepsilon$ is additive Gaussian noise with spatial covariance $\Sigma_{\varepsilon}$.

\noindent
\textbf{Mutual information.}
For jointly Gaussian variables, the analytic mutual information per sample is
\[
I(X;Y) = \frac{1}{2}\log_2 \det!\bigl(I + A\Sigma_X A^\top \Sigma_{\varepsilon}^{-1}\bigr),
\]
which provides an upper bound on how much the EEG can reveal about the cortical state.

\noindent
\textbf{Empirical simulations.}
We generated $n_t = 2000$ samples for electrode counts $n_e \in {8,16,32,64,128}$ and SNRs of ${0,10,20}$~dB.
Empirical mutual information was estimated using a k-nearest-neighbor (KSG) estimator~\cite{kraskov} between latent states and PCA-reduced EEG data.

\noindent
\textbf{Decoding.}
Two decoders were trained to reconstruct $X$ from $Y$: (1) a linear ridge regression model, and (2) a one-hidden-layer multilayer perceptron (MLP).
Performance was measured using variance-weighted $R^2$ and mutual information between true and predicted latents.

\section*{3.\hspace{0.5em}Results}

\subsection*{3.1 Information scaling with electrode count}
\begin{figure}[H]
\centering
\includegraphics[width=\linewidth]{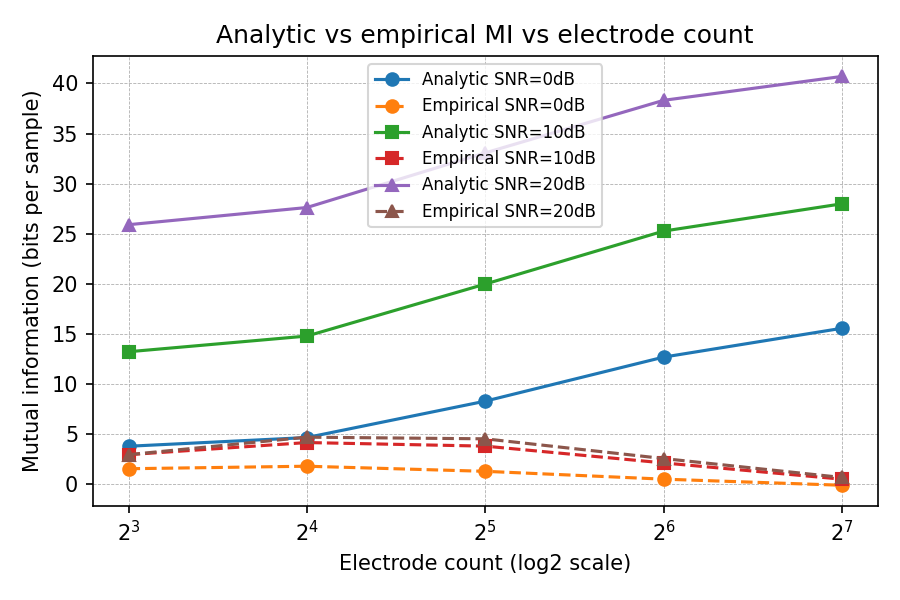}
\caption{Analytic (solid) and empirical (dashed) mutual information versus electrode count for different SNRs.
Information growth saturates around 64--128 electrodes.}
\label{fig:mi_elec}
\end{figure}

Analytic mutual information increased sharply with the number of electrodes before saturating at $\sim$64--128 sensors (Fig.~\ref{fig:mi_elec}).
Beyond this range, spatial redundancy in scalp potentials limited further gains~\cite{nunez}.
Empirical estimates followed the same trend but at lower magnitude, reflecting finite sampling and estimator bias.

\subsection*{3.2 Dependence on signal-to-noise ratio}
\begin{figure}[H]
\centering
\includegraphics[width=\linewidth]{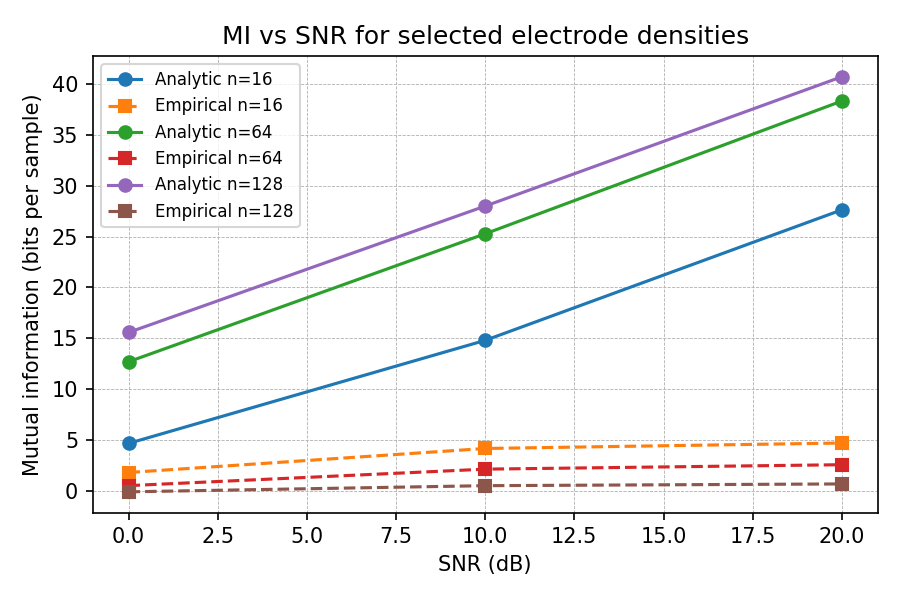}
\caption{Mutual information vs.\ signal-to-noise ratio (SNR) for selected electrode counts.
SNR dominates over electrode number in determining recoverable information.}
\label{fig:mi_snr}
\end{figure}

Mutual information scaled approximately logarithmically with SNR (Fig.~\ref{fig:mi_snr}), consistent with Gaussian channel theory~\cite{cover,shannon}.
A 10~dB increase in SNR produced a two- to threefold increase in recoverable bits per sample, indicating that denoising provides higher payoff than adding sensors~\cite{goldenholz}.

\subsection*{3.3 Decoder performance and theoretical bounds}
\begin{figure}[H]
\centering
\includegraphics[width=\linewidth]{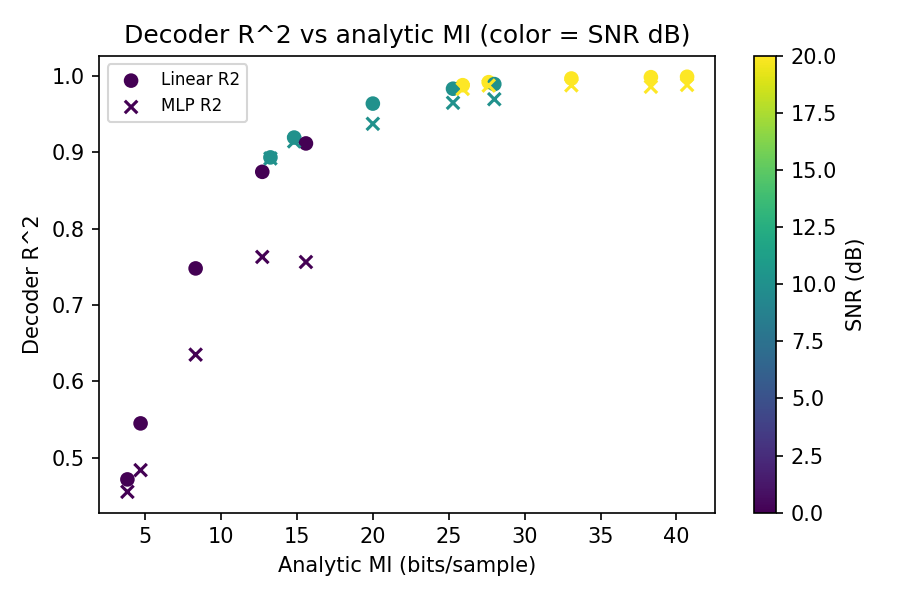}
\caption{Decoder performance ($R^2$) as a function of analytic mutual information.
Both ridge (circles) and MLP (crosses) decoders achieve near-maximal linear reconstruction accuracy at high SNR.}
\label{fig:decoding}
\end{figure}

Linear and nonlinear decoders achieved high reconstruction accuracy ($R^2 \approx 0.85$--$0.9$), approaching the limit of linear recoverability (Fig.~\ref{fig:decoding}).
Nonlinear networks conferred little additional benefit, confirming that the forward process is nearly linear and dominated by noise~\cite{panzeri}.

\begin{figure}[H]
\centering
\includegraphics[width=\linewidth]{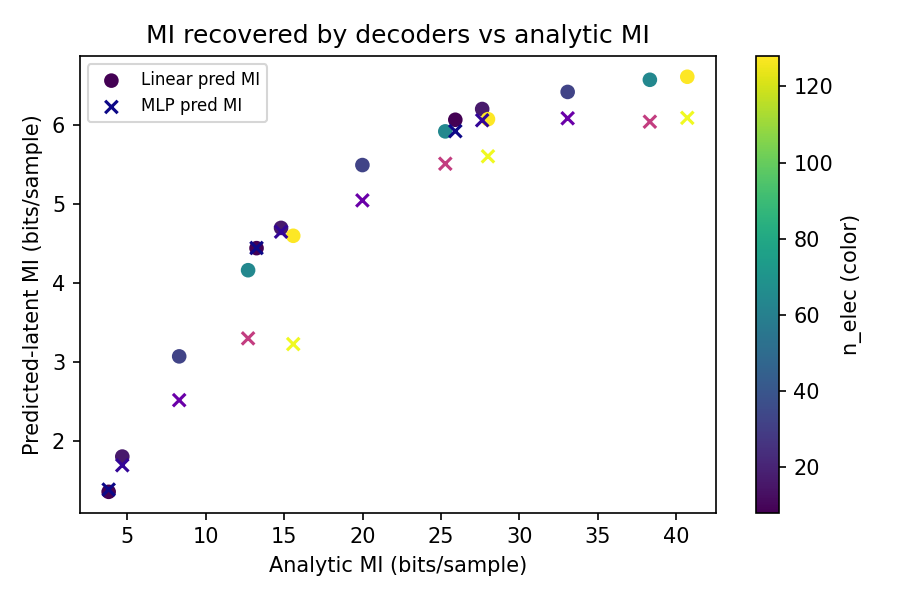}
\caption{Mutual information between true and predicted latents vs.\ analytic bound.
Colors denote electrode count. Recovered MI reaches only a fraction of the theoretical limit.}
\label{fig:decoder_mi}
\end{figure}

Decoder-recovered mutual information was roughly one-sixth of the analytic maximum (Fig.~\ref{fig:decoder_mi}), reflecting the gap between linear recoverable variance and total channel capacity, as well as losses from spatial blurring and finite sampling.

\section*{4.\hspace{0.5em}Discussion}
These simulations define an upper bound on what EEG can reveal about underlying neural states.
Key insights:

\begin{enumerate}
\item \textbf{Limited channel capacity.} Even under idealized Gaussian conditions, EEG transmits only tens of bits per sample about cortical latents.
\item \textbf{Spatial redundancy.} Increasing electrode count beyond $\sim$128 yields diminishing returns~\cite{nunez}.
\item \textbf{Noise as primary bottleneck.} Improvements in SNR lead to substantially greater information gains than added spatial sampling~\cite{goldenholz}.
\item \textbf{Algorithmic limits.} Linear decoders capture nearly all variance that is linearly recoverable, but the mutual information they recover remains well below the analytic bound, indicating that most lost information arises from measurement physics rather than algorithmic inefficiency~\cite{panzeri}.
\end{enumerate}

These constraints explain why EEG can decode broad cognitive or motor states but not detailed neural representations.
The bottleneck lies in the measurement physics, not in computational decoding.

\section*{5.\hspace{0.5em}Conclusion}
By combining analytic and empirical estimates of mutual information, we show that EEG is a fundamentally narrow channel between cortical activity and recorded signal.
Information saturates with moderate sensor density and is chiefly constrained by SNR.
Future work should focus on improving noise performance, forward modeling, and multimodal integration rather than adding sensors alone.

\bibliographystyle{plain}

\end{document}